\documentclass[10pt,conference]{IEEEtran}
\IEEEoverridecommandlockouts
\usepackage{cite}
\usepackage{amsmath,amssymb,amsfonts}
\usepackage{algorithmic}
\usepackage{graphicx}
\usepackage{textcomp}
\usepackage{hyperref}
\usepackage{xcolor}
\usepackage{xspace}
\usepackage{fancyhdr}
\newcommand{\codeseq}{\textsc{code2seq}\xspace}
\def\BibTeX{{\rm B\kern-.05em{\sc i\kern-.025em b}\kern-.08em
    T\kern-.1667em\lower.7ex\hbox{E}\kern-.125emX}}
\begin{document}

\title{CodeLens: An Interactive Tool for Visualizing Code Representations
\thanks{This work is funded by the European Union's Horizon Research and Innovation Programme under Grant Agreement n$^\circ$ 101070303.}
}


\author{\IEEEauthorblockN{1\textsuperscript{st} Yuejun Guo}
\IEEEauthorblockA{IT for Innovative Services (ITIS) \\
Luxembourg Institute of Science and Technology (LIST)\\
Luxembourg \\
yuejun.guo@list.lu}
\and
\IEEEauthorblockN{2\textsuperscript{nd} Seifeddine Bettaieb}
\IEEEauthorblockA{IT for Innovative Services (ITIS) \\
Luxembourg Institute of Science and Technology (LIST)\\
Luxembourg \\
seifeddine.bettaieb@list.lu}
\and
\IEEEauthorblockN{3\textsuperscript{rd} Qiang Hu}
\IEEEauthorblockA{SnT \\
University of Luxembourg\\
Luxembourg \\
qiang.hu@uni.lu}
\and
\IEEEauthorblockN{4\textsuperscript{th} Yves Le Traon}
\IEEEauthorblockA{SnT \\
University of Luxembourg\\
Luxembourg \\
yves.letraon@uni.lu}
\and
\IEEEauthorblockN{5\textsuperscript{th} Qiang Tang}
\IEEEauthorblockA{IT for Innovative Services (ITIS) \\
Luxembourg Institute of Science and Technology (LIST)\\
Luxembourg \\
qiang.tang@list.lu}
}

\maketitle
\thispagestyle{fancy}
\pagestyle{fancy}
\cfoot{\thepage}
\renewcommand{\headrulewidth}{0pt} 
\renewcommand{\footrulewidth}{0pt}

\begin{abstract}
Representing source code in a generic input format is crucial to  automate software engineering tasks, e.g., applying machine learning algorithms to extract information. Visualizing code representations can further enable human experts to gain an intuitive insight into the code. Unfortunately, as of today, there is no universal tool that can simultaneously visualise different types of code representations. In this paper, we introduce a tool, CodeLens, which provides a visual interaction environment that supports various representation methods and helps developers understand and explore them. CodeLens is designed to support multiple programming languages, such as Java, Python, and JavaScript, and four types of code representations, including sequence of tokens, abstract syntax tree (AST), data flow graph (DFG), and control flow graph (CFG). By using CodeLens, developers can quickly visualize the specific code representation and also obtain the represented inputs for models of code.  The Web-based interface of CodeLens is available at \url{http://www.codelens.org/}. The demonstration video can be found at \url{http://www.codelens.org/demo}.
\end{abstract}

\begin{IEEEkeywords}
code representation, interactive visualization
\end{IEEEkeywords}

\section{Introduction}
\label{sec:introduction}
The development of machine learning (ML) has enabled the automation of a wide range of Software Engineering (SE) tasks~\cite{codexglue2021paper}. For example, large language models (LLMs), e.g., CodeBERT~\cite{codebert2020} and CodeX~\cite{codex2021openai}, have proven to achieve the state-of-the-art performance in clone detection, vulnerability detection, and code generation. One prerequisite to using ML models is the code representation~\cite{Samoaa2022mapping}, which involves the transformation of source code into analyzable data by the ML models.


In the domain of SE, different code representations have been developed and studied, such as the text-based sequence of tokens~\cite{codebert2020}, tree-based abstract syntax tree (AST)~\cite{code2vec2019,alon2018codeseq}, graph-based data flow graph (DFG)~\cite{graphcodebert2020,graphsearch2021liu} and control flow graph (CFG)~\cite{aho2007compilers}. Before adopting code representations for downstream tasks, human experts often need to interpret them so that proper actions can be taken. Without visualization, interpretation is a challenging task even for experienced developers. In addition, when using these representations (e.g., AST and CFG) in ML models, complex parsing libraries and toolkits are required to be installed to process source code. For example, to convert source code to the AST format, there are the tree-sitter~\cite{treesitter2023} for multiple programming languages, Joern~\cite{joern} for C/C++, and Python parser~\cite{pythonparser} for Python. Installing and understanding all these libraries can be time-consuming and complex, as a result, the whole process of code learning can not be easily automated. An additional challenge is that there is no library for some formats such as DFG and CFG.

\begin{table*}[ht]
\caption{Comparison between CodeLens and existing code visualization tools.}
\label{tab:comparison}
\resizebox{1.\textwidth}{!}{
\begin{tabular}{lcccc}
\hline
\textbf{Tool name} & \textbf{Support language} & \textbf{Support representation} & \textbf{Visualization type} & \textbf{Download type} \\ \hline
AST Visualization on browser~\cite{browser2012} & JavaScript & AST & Graph & Image, JSON \\
Java Parser~\cite{javaparser2016} & Java & AST & - & JSON \\
\codeseq~\cite{alon2018codeseq} & Java & AST & Graph & Image \\
Code2flow~\cite{code2flow} & Unknown & - & Graph & Image \\
Swift AST Explorer~\cite{swiftastexplore} & Swift & AST & Text & - \\
JavaScript AST Visualiser~\cite{jsvis} & JavaScript & Sequence of tokens, AST & Text & Image \\
code2vec~\cite{code2vec2019} & Java & AST & Graph & Image \\ \hline
CodeLens & \begin{tabular}[c]{@{}c@{}}Java, Python,\\ JavaScript\end{tabular} & \begin{tabular}[c]{@{}c@{}}Sequence of tokens, AST,\\ DFG, CFG\end{tabular} & Text, graph & Image, JSON \\ \hline
\end{tabular}
}
\end{table*}

In the literature, several tools have been proposed to visualize code representations and help developers understand code. A comparison of them is shown in Table~\ref{tab:comparison}. Among them, those from~\cite{browser2012, alon2018codeseq, code2flow} receive source code as input and then visualize it as an AST graph. Such graphs can be  downloaded for further analysis. Besides, the Java Parser from~\cite{javaparser2016} can transfer Java code to AST with JSON format for understanding Java programs. Other tools such as~\cite{swiftastexplore, jsvis} can transfer and visualize programs as the text of tokens or ASTs. However, there are a few limitations w.r.t. the existing tools: 1) they can only support one specific programming language (e.g., Java or JavaScript), 2) they only support one visualization type~(e.g., graph or text), and 3) after analyzing code, only one format of data can be downloaded~(image or JSON). These issues seriously limit the usage of these tools in practical deployment.  

To address these limitations, in this paper, we introduce CodeLens, an interactive tool for visualizing different code representations for different programming languages. In more detail, CodeLens supports three popular programming languages, including Java, Python, and JavaScript. It supports four types of code representation, including sequence of tokens, AST, DFG, and CFG. To our knowledge, this is the first tool that supports visualization of both DFG and CFG. Furthermore, CodeLens supports two types of visualization, i.e., text and graph, and provides two formats of outputs, i.e., image data and JSON files, which save all the represented code information and can be downloaded for further usage. In the evaluation part, we demonstrate the usefulness of CodeLens with two straightforward use cases, 1) visualizing different types of code representations to help users understand code intuitively, and 2) providing different types of pre-processed inputs for machine learning models. It is worth noting that CodeLens can be used in many other use cases by providing the necessary input for different code analysis tasks. Due to the space limitation, we omit the details in this paper. 



\section{C\lowercase{ode}V\lowercase{is} Overview}

\label{sec:overview}
\begin{figure}[htpb]
    \centering
    \includegraphics[scale=0.322]{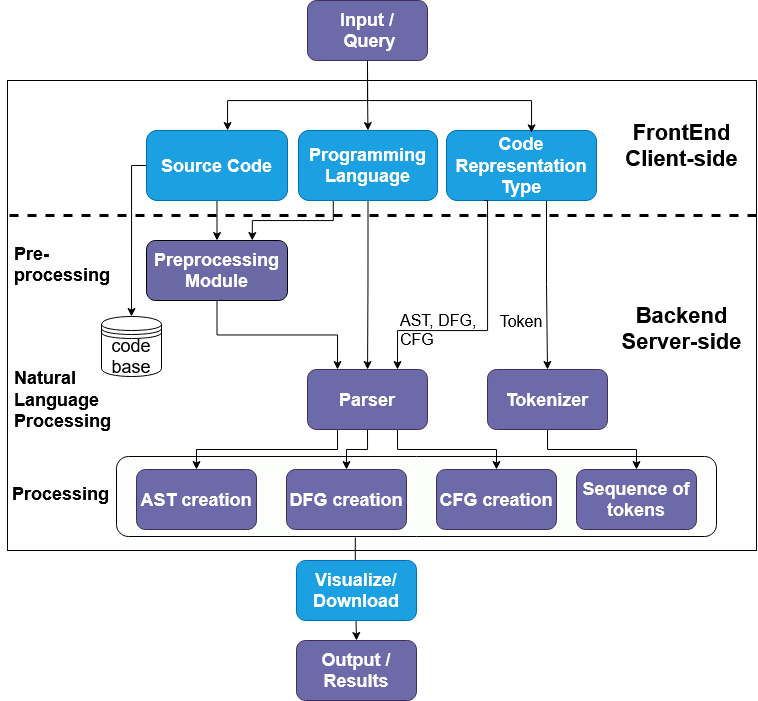}
    \caption{Architecture of CodeLens.}
    \label{fig:overview}
\end{figure}

Figure~\ref{fig:overview} provides a comprehensive overview of the CodeLens architecture, which consists of a Web-based frontend (written in JavaScript using React) and a server-side backend (written in Python) connected through the Flask microframework. The frontend (client) provides a user-friendly interface, allowing developers and users to interact with different code representations seamlessly. The backend (server) of CodeLens plays a crucial role in processing code and generating graphical code representations of different formats.

\subsection{User Interface}
\label{subsec:interface}
The user interface prioritizes simplicity, incorporating two distinct boxes. On the left-hand side, users are provided with a console-like interactive environment where they can input their code. To cater to the diverse programming preferences of users, our interface offers a selection of three programming languages: Python, JavaScript, and Java. Within each language category, users can explore and experiment with five different code examples, providing them ample opportunities for practice. On the right-hand side, users can visualize or download the resulting output, ensuring a comprehensive understanding of the code's execution.

\begin{figure}[htpb]
    \centering
     \includegraphics[scale=0.175]{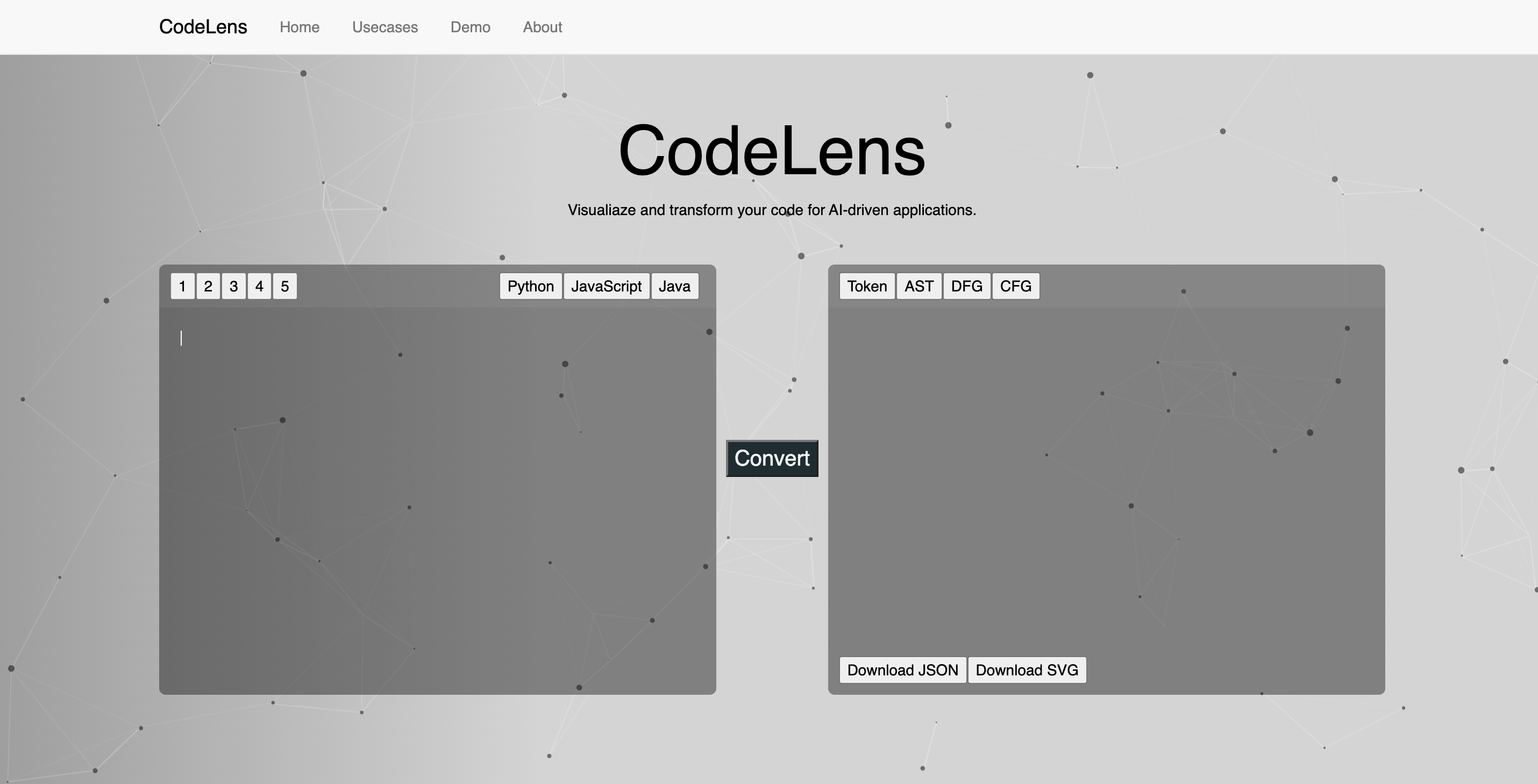}
    \caption{User interface design of CodeLens.}
    \label{fig:interface}
\end{figure}

To facilitate efficient processing, a {\bf Convert} button is positioned between the two boxes, enabling users to initiate the transformation of their input. For a visual depiction of this intuitive interface, please refer to Figure~\ref{fig:interface}.

\subsection{Back-End Implementation}
\label{subsec:backend}
The backend of CodeLens implements the transformations from source code (i.e., Java, Python, and JavaScript) to code representations (i.e., text-based sequence of tokens, AST, DFG, CFG). 

\paragraph{Sequence of tokens} A source code is treated as plain text and processed into a linear sequence of tokens via a tokenizer. Each line of code is chopped into pieces by looking for the whitespace (tabs, spaces, newlines). Each piece is finally represented by an integer that refers to the ID of the piece in a so-called vocabulary. A piece can be a word, a subword, or a character depending on different tokenizers~\cite{token2021nlp}. The subword-based tokenizer, Byte-Pair Encoding (BPE)~\cite{BPE1994gage} is implemented in CodeLens due to its popularity in code-related DL models. 

\paragraph{Abstract syntax tree (AST)} An AST is a tree representation of the abstract syntactic structure of a piece of source code. Each node in the tree represents a construct occurring in the source code. When converting a piece of source code to an AST, only structural information is preserved, such as variable types, order and definition of executable statements, and identifiers. In CodeLens, the tree-sitter\cite{treesitter2023} library is used to parse source code in different programming languages.

\paragraph{Data flow graph (DFG)} As the name suggests, DFG~\cite{graphcodebert2020} is a data-oriented graph representation that shows the flow of data through a piece of source code. In a DFG, each node represents a variable or an expression, and each edge represents the flow of data between them. In CodeLens, the DFG is extracted from the AST of the given source code by tracing the variable or expression statement according to the programming grammar in the underlying programming language. 

\paragraph{Control flow graph (CFG)} Similar to DFG, CFG~\cite{aho2007compilers} is a graph-based representation. While CFG is process-oriented, it represents all paths that might be traversed through the execution. In a CFG, nodes portray basic blocks of statements and conditions, and edges describe the transfer of control and subsequent access or modification onto the same variable. For instance, a for-loop is a basic control flow statement for specifying iteration. Note that, a CFG includes two designated blocks, an entry block and an exit block where the control enters and leaves the flow. In CodeLens, the CFG is extracted from the AST of the given source code by tracing all control statements defined in the underlying programming language.

\section{Use Cases}
\label{sec:usage}
In this section, we present two use cases to demonstrate the usage of CodeLens.

\begin{figure}[h]
    \centering
    \includegraphics[width=0.5\textwidth]{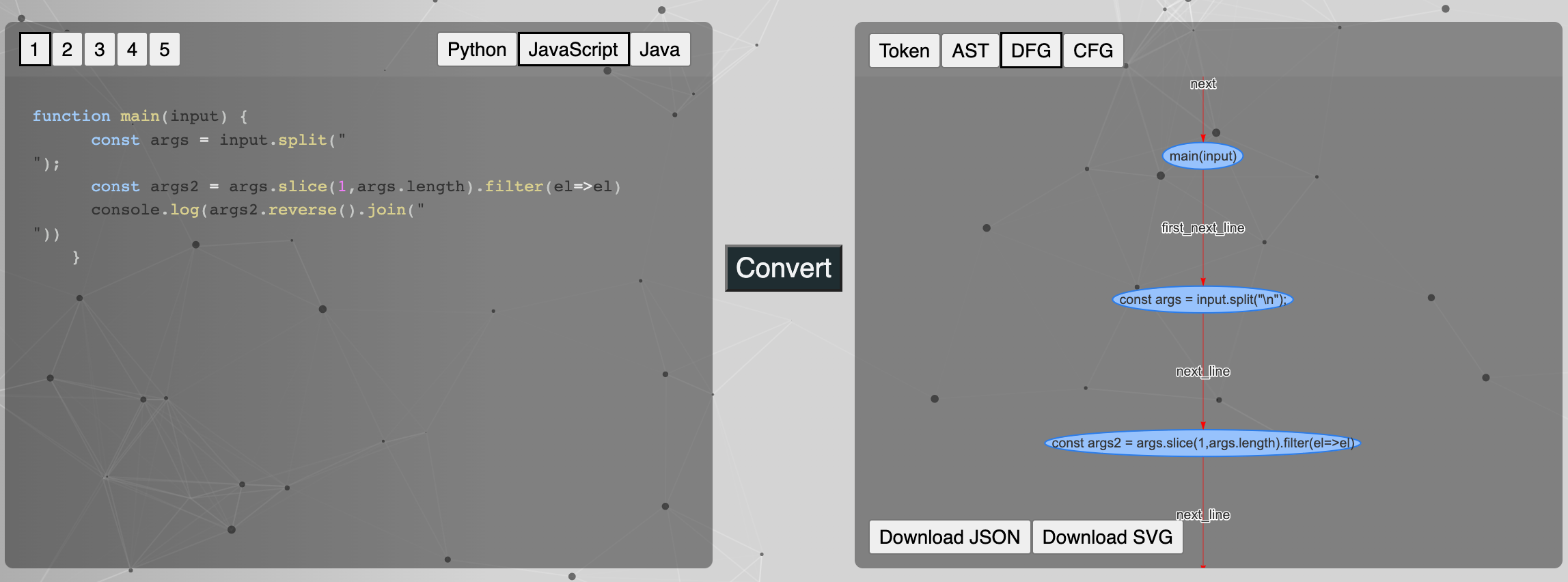}
    \caption{Example of a Java DFG visualized by CodeLens.}
    \label{fig:cfg}
\end{figure}

\paragraph{Code representation visualization} The first usage of CodeLens is to visualize code representation. CodeLens receives raw code snippets and then draws the code representations in the right panel as shown in Figure~\ref{fig:cfg}. The figures can be downloaded into different formats for further analysis. For example, a user can use them to check the change of representation after the change of code snippets or to compare two representations of two programs.

\begin{figure}[h]
    \centering
    \includegraphics[width=0.48\textwidth]{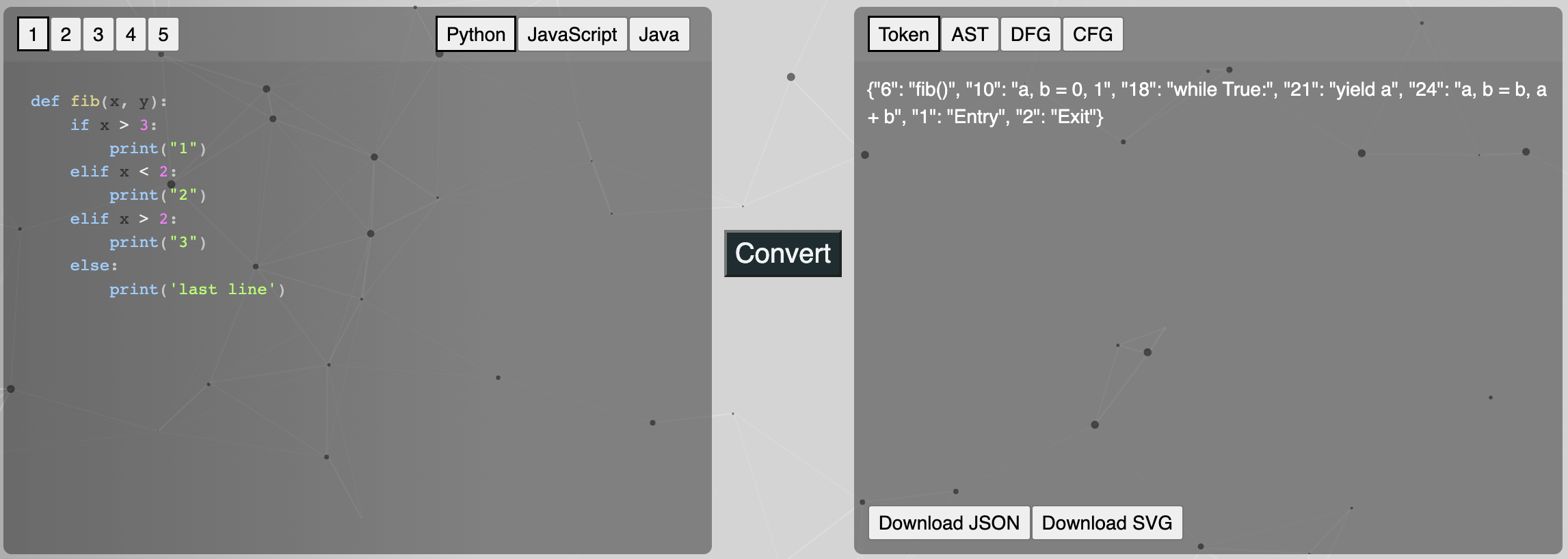}
    \caption{An example of transformed code data that can be used for tokenizer.}
    \label{fig:tokens}
\end{figure}

\paragraph{Inputs to ML models extraction} In addition to representation visualization, CodeLens also supports producing pre-processed inputs for ML models. This means that users can easily test their machine-learning models of code without preparing other packages like Java/Python Parser. As shown in Figure~\ref{fig:tokens}, the transformed code data (e.g., tokens, AST) can be downloaded and fed to the tokenizer provided by pre-trained code models, for example, CodeBERT~\cite{codebert2020}.  



\section{Related Work}
\label{sec:related}
\paragraph{Code representations in ML models}
Machine learning (ML), particularly deep learning (DL), models have been proven successful in automating various software engineering tasks, such as problem classification~\cite{hu2023codes}, clone detection~\cite{zubkov2022evaluation}, and vulnerability detection~\cite{zhou2019devign}. Code representation is a preliminary step to convert source code into a readable format for all models. Among existing code representations, the representation of a sequence of tokens is mostly used in large language models (LLMs) for code that supports multiple downstream tasks. For example, CodeBERT~\cite{codebert2020}, CodeT5~\cite{codet52021wang}, and CodeX~\cite{codex2021openai} take the sequence of tokens of source code by the BPE tokenizer as input. In addition to the sequence of tokens by the BPE tokenizer, GraphCodeBERT~\cite{graphcodebert2020} adds the data flow graph (DFG) representation of source code to capture the relation of ``where-the-value-comes-from'' between variables. The task-specific GraphSearchNet~\cite{graphsearch2021liu} model also considers the DFG representation of source code to undertake the code search task. \codeseq~\cite{alon2018codeseq} and code2vec~\cite{code2vec2019} extract a set of syntactic paths from the AST representation of source code. 

\paragraph{Online tools for code visualization} Multiple tools with Web-based interface are available, such as the AST Visualization on browser~\cite{browser2012} and Java Parser~\cite{javaparser2016}. However, as shown in the comparison in Table~\ref{tab:comparison}, these tools have four main limitations. First, only one programming language is supported. For example, Java Parser~\cite{javaparser2016}, \codeseq~\cite{alon2018codeseq}, and code2vec~\cite{code2vec2019} only support Java. Second, the support code representation mainly focuses on AST. None of the seven tools support graph-based representation. Third, the visualization type is limited. Visualizing the nodes and edges of an AST in a tree diagram is intuitive but challenging. The Swift AST Explorer~\cite{swiftastexplore} and JavaScript AST Visualiser~\cite{jsvis} simply visualize an AST in a tree structure, which is less helpful to understand the source code. Last, the output of most tools is limited to images. If a developer is willing to use the representation, e.g., AST generated by a tool, a JSON file recording the nodes and edges should be provided. In contrast, CodeLens has addressed all four limitations.

\section{Conclusion}
\label{sec:conclusion}
In this paper, we have introduced an interactive tool named CodeLens, which visualizes four most popular code representations widely used by machine learning (ML) models, including sequence of tokens, AST, DFG, and CFG. The tool allows software and application developers to understand how source code is represented when applying ML models to automate software engineering tasks. In addition, it can also serve as an interface between source code and numeral code representations in ML models. This eliminates the necessity for developers to install complex parsing libraries and toolkits, through making the code representation process more user-friendly. 

CodeLens opens many directions for future work. For example, CodeLens can be extended to support more programming languages (e.g., C, C++ and PHP) and other useful code representations. Another direction is to develop applications on the basis of CodeLens, e.g., integrating a vulnerability detection module into CodeLens to perform vulnerability detection tasks without the need to pre-processing the code in advance. 


\section*{Acknowledgment}
An experienced developer Bowen Liu (who is currently working on ML-based malware detection) was invited to test CodeLens. We thank him for his valuable feedback.

\bibliographystyle{IEEEtran}
\bibliography{IEEEabrv, refs}

\end{document}